\documentclass[aps,prc,preprint,groupedaddress,showpacs]{revtex4}
\usepackage{graphicx}
\usepackage{dcolumn}
\usepackage{bm}

\begin{document}

\preprint{IKDA 04/13, SFB 634 04/13}

\title{Systematic study of ($\gamma$,n) reaction rates 
for $Z \geq 78$ isotopes}


\author{K.\ Sonnabend}
\author{K.\ Vogt}
\author{D.\ Galaviz}
\author{S.\ M\"uller}
\author{A.\ Zilges}
\email{zilges@ikp.tu-darmstadt.de}
\affiliation{Institut f\"ur Kernphysik, Technische Universit\"at Darmstadt, Schlossgartenstr.\ 9, D--64289 Darmstadt, Germany}


\date{\today}

\begin{abstract}
The ($\gamma$,n) reaction rates of the isotopes 
$^{196,198,204}$Hg and $^{204}$Pb have been 
determined using the photoactivation technique 
in an energy region relevant for $p$ process 
nucleosynthesis. The systematic study of the 
ground-state ($\gamma$,n) reaction rates on 
even-even nuclei in the mass region $Z \geq 78$ is 
complemented with these experiments. The data are 
compared to rates predicted in the framework of 
two statistical model approaches.
\end{abstract}


\maketitle


\section{Introduction\label{sec:intro}}
The heavy nuclei with $Z \geq 26$ are mainly produced 
by the astrophysical $s$ and $r$ process. Both describe 
nucleosynthesis by neutron capture reactions with subsequent 
$\beta$ decays. The $s$ process takes place during quiescent 
burning phases of medium mass stars (mean neutron 
density $n_{\rm n} \approx 2 - 4 \times 10^{8}\ {\rm cm}^{-3}$, 
mean temperature $kT \approx 25\ {\rm keV}$ \cite{kaep99}). 
In contrast, the $r$ process requires explosive 
environments ($n_{\rm n} \approx 10^{20}\ {\rm cm}^{-3}$, 
$T \approx 3 \times 10^{9}\ {\rm K}$ \cite{wall97}).

However, 35 stable isotopes on the proton rich side 
of the valley of stability cannot be produced by 
either of these processes. A complete list of 
the so-called $p$ nuclei can be found in \cite{lamb92,arno03}. 
Their natural abundances are in the order of 0.01\% 
to 1\%, a hint for their production in a secondary 
process. The only exceptions from the low abundances
are $^{92,94}$Mo and $^{96}$Ru.
One has to distinguish between the $p$ process 
in the mass region $Z \leq 50$ 
(e.g.\ $rp$ process \cite{thie01,scha01}) and 
at higher masses where photodisintegration 
reactions like ($\gamma$,n), ($\gamma$,$\alpha$), 
and ($\gamma$,p) play the important role.

The latter reactions take place at temperatures 
of $T = 2 - 3 \times 10^{9}\ {\rm K}$ and 
the whole process lasts in the order of seconds. 
Possible astrophysical sites for this process 
are the oxygen- and neon-rich layers of type II 
supernovae. However, a definite conclusion is 
still missing. Details of the reaction path sometimes 
denoted as $\gamma$ process can be found in 
various reviews \cite{woos78,lamb92,wall97,arno99,lang99,arno03}.

The reaction network of the $p$ process is 
very extensive dealing with around 2000 nuclei 
and several thousand corresponding reaction rates. 
Thus, it is mandatory to use theoretical 
predictions because many of the nuclei involved 
are not accessible with the present experimental 
methods. However, it was recently emphasized by 
Arnould and Goriely \cite{arno03} that the present 
lack of measured reaction rates in the astrophysically 
relevant energy region is a constraint on the 
reliability of theoretical predictions. Most 
of the existing experimental data on photodisintegration 
rates was measured around the Giant Dipole Resonance, 
therefore, being far off the energy region of interest 
for $p$ process nucleosynthesis lying close above the 
reaction threshold.  

The ($\gamma$,n) reaction rates of the most proton rich stable
isotopes can be measured by photoactivation. High resolution
$\gamma$ spectroscopy of transitions in the daughter nuclei of the
produced unstable isotopes allows a very high sensitivity.
However, the experiments are sometimes hampered by high neutron
separation energies $S_{\rm n}$ and very low abundances of the
isotopes of interest.

In this manuscript we present the results of a systematic
investigation of even-even neutron rich isotopes with $Z \geq 78$.
The experimental method is presented in Section~\ref{sec:expMeth}
followed by a description of two different ways to evaluate the
data. Section~\ref{sec:result} summarizes the results for the
observed Hg and Pb isotopes. A comparison to different theoretical
predictions is drawn in Section~\ref{sec:theory} including
previous results on $^{190,192,198}$Pt and $^{197}$Au. We conclude
with a summary and outlook.

\section{Experimental method\label{sec:expMeth}} The experiments
were performed at the superconducting Darmstadt electron
accelerator S--DALINAC \cite{rich96}. The monoenergetic electron
beam is fully stopped in a thick copper radiator target. Thus, a
continuous bremsstrahlung spectrum is produced with energies up to
the electron energy $E_{\rm max}$. The targets are usually placed
behind a collimator made of copper to get a well defined beam
spot. The absolute intensity of the photon beam is determined by
an online measurement of the reaction $^{11}$B($\gamma$,$\gamma
'$) using two high-purity germanium detectors. The energy
distribution results from a Monte Carlo simulation that is fitted
to these data at several energies $E_{\rm max}$. Details of the
setup are described in \cite{mohr99a,hart00,vogt01b}.

If a higher photon intensity is needed, e.g.\ due to a low amount
of target material, the targets are positioned directly behind the
radiator target where the intensity of the beam is about a factor
of 300 higher. Here the determination of the absolute photon
intensity is realized by measuring relative to a standard
reaction. Either the reaction $^{197}$Au($\gamma$,n) or
$^{187}$Re($\gamma$,n) is used. The cross sections of both
reactions are well known close above their respective reaction
thresholds \cite{vogt02,muel04b}.

The targets are typically irradiated between 12 and 24 hours
depending on the expected activation rate. Afterwards, the yield
of the produced unstable isotopes is measured offline. For this
purpose the $\gamma$ rays emitted after the $\beta$ decay of the
unstable nuclei are detected. 

The targets are mounted in front of a well shielded HPGe
detector. The number of $\gamma$ rays $Y$ is directly proportional
to the integrated product of the ($\gamma$,n) cross section
$\sigma (E)$ and the photon flux $N_{\gamma}(E,E_{\rm max})$, i.e.:
\begin{equation}\label{eq:yield} Y \propto \int_{0}^{\infty}
N_{\gamma}(E,E_{\rm max}) \sigma (E) dE \end{equation}
The factor of proportionality depends on the activation time as
well as on the absolute detector efficiency and the absolute
intensity of the observed $\gamma$ decay line. For a detailed
discussion see Ref.~\cite{vogt01b}.

Due to the high sensitivity of the photoactivation technique one
can use naturally composed target material in many cases. Metallic
discs were used for the investigation of Pb whereas for the
observation of the Hg isotopes the targets were composed of HgS
powder that was pressed into thin tablets. The properties of the
targets as well as of the calibration standards used are listed in
Table~\ref{tab:target}.

\begin{table} \caption{\label{tab:target}Properties of the targets
and calibration standards used for the photoactivation
experiments.} \begin{ruledtabular} \begin{tabular}{cccc} target
& masses          & obs.\ isotopes & nat.\ abundances
\cite{biev93} \\ \hline $^{\rm nat}$Pb  & 435 -- 443\ mg  &
$^{204}$Pb    & (1.4~$\pm$~0.1)\%   \\ \hline &                 &
$^{196}$Hg    & (0.15~$\pm$~0.01)\% \\ $^{\rm nat}$HgS       &
1.97 -- 2.72\ g & $^{198}$Hg    & (9.97~$\pm$~0.08)\% \\ &
& $^{204}$Hg    & (6.87~$\pm$~0.04)\% \\ \hline $^{\rm nat}$B   &
636 -- 846\ mg  & $^{11}$B      & (80.1~$\pm$~0.2)\%  \\ \hline
$^{\rm nat}$Au  & 151 -- 165\ mg  & $^{197}$Au    & 100\%
\end{tabular} \end{ruledtabular} \end{table}

The Pb targets were activated behind the collimator using $^{11}$B
as well as $^{197}$Au to determine the photon beam intensity.
Thus, a disc of Au and one of Pb were sandwiched between two thin
layers of boron. Due to the low abundance of $^{196}$Hg, the Hg
targets were placed in the more intense photon flux directly
behind the radiator target. They were mounted between two thin
foils of Au to calibrate the photon beam intensity.

\section{Data analysis \label{sec:analysis}}
The ($\gamma$,n) reaction rate $\lambda (T)$ for a nucleus in a
thermal photon bath at a certain temperature $T$ is given by
\begin{equation}\label{eq:lambda} \lambda (T) = \int_{0}^{\infty}
c  n_{\gamma}(E,T) \sigma (E) dE \end{equation}
where $c$ is the speed of light and $\sigma (E)$ is the cross
section of
the ($\gamma$,n) reaction. The number of photons with energy $E$
per unit volume and energy interval $n_{\gamma}(E,T)$ is described
by the Planck distribution \begin{equation}\label{eq:planck}
n_{\gamma}(E,T) = \left( \frac{1}{\pi} \right)^{2}  \left(
\frac{1}{\hbar c} \right)^{3}  \frac{E^{2}}{\exp(E/kT) - 1}
\end{equation} In order to determine the reaction rate $\lambda
(T)$ at a given temperature $T$ Eq.~\ref{eq:lambda} offers two
possibilities: one can either derive the energy dependence of the
cross section $\sigma (E)$ and calculate $\lambda (T)$ using
Eqs.~\ref{eq:lambda} and \ref{eq:planck} or measure the
reaction rate $\lambda (T)$
directly by approximating the Planck distribution of photons at a
given temperature in the astrophysically relevant energy range.
The former case is the {\it conventional} method
whereas the latter one will be called the
{\it superposition} method. The pros and cons
of the two methods are explained in the following paragraphs.

\subsection{The superposition method\label{subsec:super}} 
It is very userful to derive the experimental
reaction rates without any assumption about the 
shape of the cross
section close to the threshold.
Therefore, the superposition method approximates the Planck
distribution of Eq.~\ref{eq:planck} by a superposition of several
bremsstrahlung spectra with different endpoint energies in the
region of astrophysical interest.

\begin{figure}
\includegraphics[width=\columnwidth]{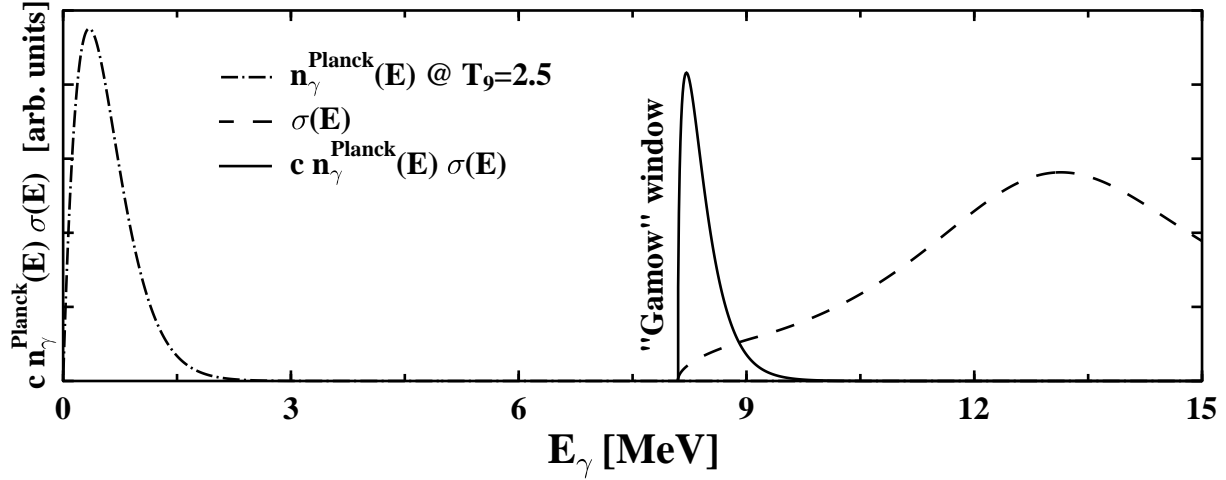}
\caption{\label{fig:gamow} The Gamow-like energy window for
($\gamma$,n) reactions. The Planck distribution corresponding to a
temperature $T = 2.5 \times 10^{9}\ {\rm K}$ is drawn
(dashed-dotted line). Note that this distribution shows an
exponential decrease. A ($\gamma$,n) cross section with the Giant
Dipole Resonance around 13\ MeV and a typical threshold behavior
is plotted (dashed line). The product of both curves -- the
integrand of Eq.~\ref{eq:lambda} -- yields a Gamow-like energy
window as known from charged particle reactions (solid line) above
the neutron separation energy $S_{\rm n}$.} \end{figure}

Figure~\ref{fig:gamow} shows the Planck distribution at $T = 2.5
\times 10^{9}\ {\rm K}$, a typical ($\gamma$,n) cross section and
the product of both at this temperature (see Eq.~\ref{eq:lambda}).
It is obvious that an approximation of the Planck distribution in
a rather narrow Gamow-like energy window above the threshold
energy is sufficient to derive the reaction rate $\lambda (T)$
without further assumptions.

\begin{figure}
\includegraphics[width=\columnwidth]{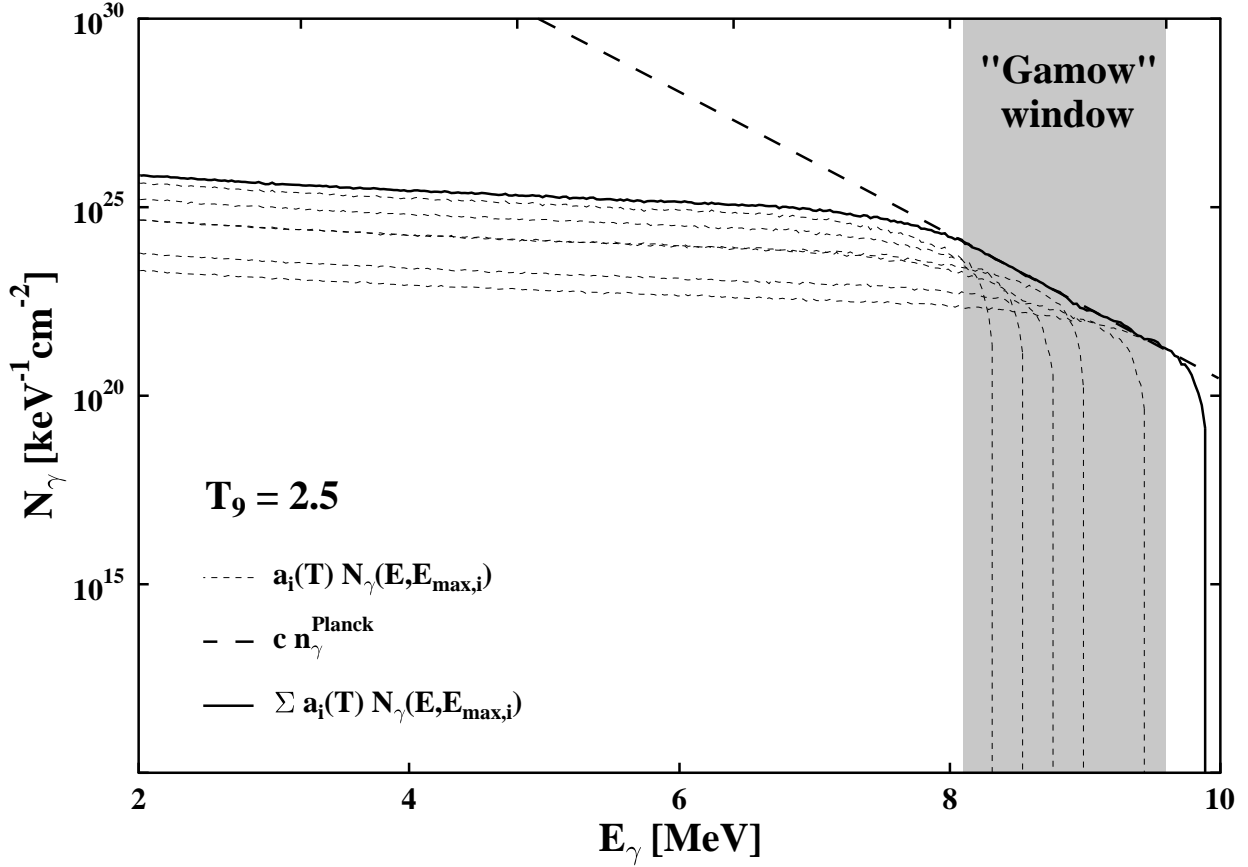}
\caption{\label{fig:approx} Approximation of the Planck
distribution by bremsstrahlung spectra. The grey shaded area
corresponds to the Gamow-like energy window i.e.\ the relevant
energy region for $p$ process nucleosynthesis. The six
bremsstrahlung spectra are weighted by temperature dependent
factors $a_{i}(T)$ (thin dashed lines). They are summed (solid
line) to approximate the Planck distribution (dashed line).}
\end{figure}

Figure~\ref{fig:approx} shows the approximation for $T = 2.5
\times 10^{9}\ {\rm K}$ using six bremsstrahlung spectra with
$E_{\rm max} \in [8325, 9900]\ {\rm keV}$. Depending on the
temperature $T$ the bremsstrahlung spectra $N_{\gamma}(E,E_{{\rm
max},i})$ have to be weighted by factors $a_{i}(T)$:
\begin{equation}\label{eq:approx} c  n_{\gamma}(E,T) \approx
\sum_{i} a_{i}(T)  N_{\gamma}(E,E_{{\rm max},i})
\end{equation} Combining Eqs.~\ref{eq:yield} and \ref{eq:approx}
with the definition of the ground-state reaction rate $\lambda
(T)$ (see Eq.~\ref{eq:lambda}) one gets a fully model independent
expression: \begin{eqnarray} \lambda (T) & \approx & \sum_{i}
a_{i}(T)  \int N_{\gamma}(E,E_{{\rm max},i})  \sigma (E)
 dE \nonumber \\ & \propto & \sum_{i} a_{i}(T) Y_{i}
\label{eq:lambdaApp} \end{eqnarray} Note that it is
possible to choose the temperature $T$ off--line 
by simply adjusting the
weighting factors $a_{i}(T)$ once the yields $Y_{i}$ have been
determined from the experimental data. It is sufficient to measure
at five to seven different energies $E_{\rm max}$ to get a good
approximation with deviation of less than about 10\% of the Planck
distribution in the astrophysically relevant energy region and for
temperatures in the range from $T = 2.0 - 3.0 \times 10^{9}\ {\rm
K}$.

The superposition method uses the high intensities of photon
fluxes that are available if bremsstrahlung spectra are taken for
the activation of the targets. Thus, it has been possible to
determine experimentally the ground-state reaction rates of rare
proton rich isotopes like $^{196}$Hg. 
points of the $p$ process reaction network.  In addition the
influence of resonances in the Gamow-like energy window is
included because of the continuous character of the photon energy
distribution.

\subsection{The conventional method\label{subsec:conv}} The setup
we have used for the activation experiments is limited to an
energy of $E_{\rm max} = 10\ {\rm MeV}$. Depending on the
investigated nucleus it is not always possible to use the
superposition method with reasonable accuracy. In these cases we
have applied the conventional method. Here we have to assume a
certain energy dependence of the ($\gamma$,n) cross section near
threshold.  \begin{equation}\label{eq:sigma} \sigma (E) =
\sigma_{0} \left( \frac{E - S_{\rm n}}{S_{\rm n}}
\right)^{0.5} \end{equation} $S_{\rm n}$ is the threshold of the
($\gamma$,n) reaction, the exponent $k = 0.5$ corresponds to a
pure $s$ wave neutron emission. The normalization factor
$\sigma_{0}$ can be derived from the experimental activation data
to calculate the reaction rate using Eq.~\ref{eq:lambda}.

Obviously, this method depends strongly on the correctness of the
assumed threshold behavior. However, a direct measurement of the
energy dependence, e.g.\ with monoenergetic photons as described
in \cite{utsu03}, is often difficult due to the large amounts of
isotopically enriched material which are needed.

\section{Experimental results\label{sec:result}}
\subsection{$^{196,198,204}$Hg \label{subsec:Hg}} 
\begin{figure}
\includegraphics[width=\columnwidth]{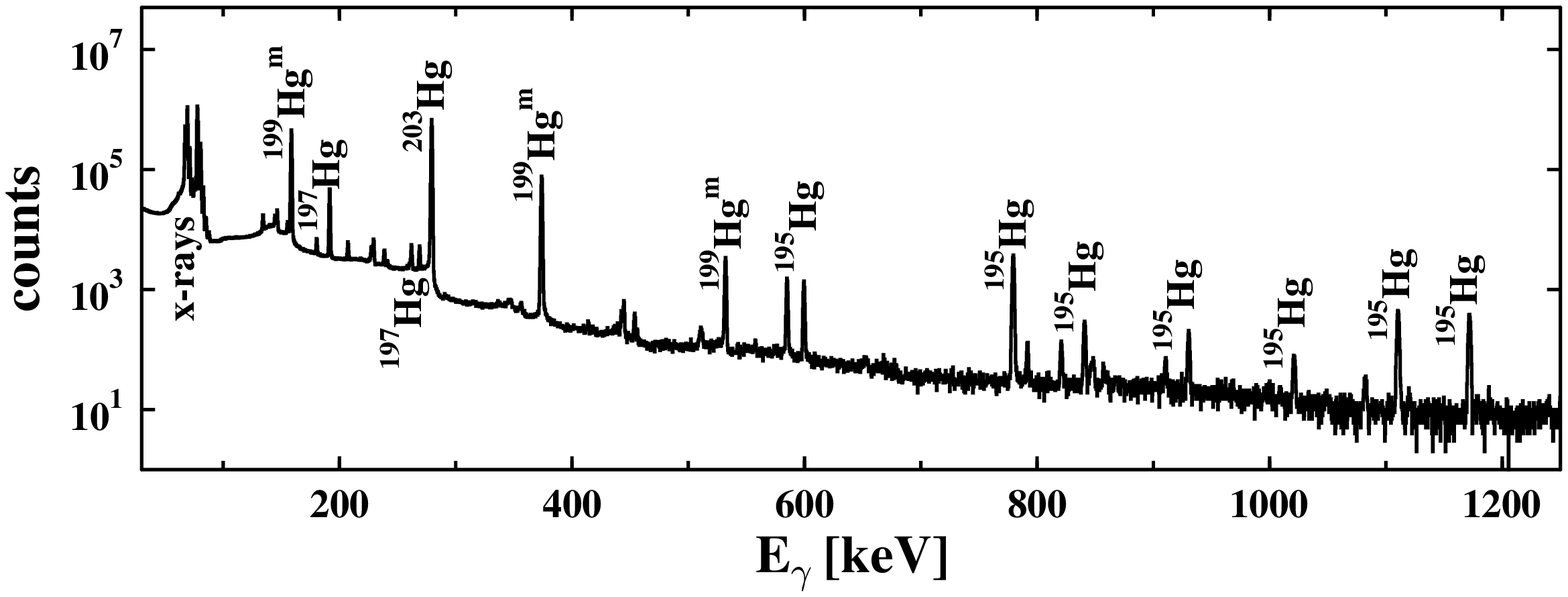} \caption{\label{fig:Hg}
Typical $\gamma$ spectrum after photoactivation of naturally
composed Hg. The spectrum was measured after an activation of
about 12\ hours with $E_{\rm max} = 9900\ {\rm keV}$. The decay
lines are indicated by the Hg isotopes produced in ($\gamma$,n) or
($\gamma$,$\gamma '$) reactions, respectively.} \end{figure}

Due to the low
natural abundance of $^{196}$Hg the $^{\rm nat}$Hg targets were
activated directly behind the radiator target to use the higher
photon intensity at this place of the experimental setup (see
Section~\ref{sec:expMeth}). The properties of the activation
reactions as well as of the decay of the produced unstable
isotopes are summarized in Table~\ref{tab:propHg}.
Figure~\ref{fig:Hg} shows a typical decay spectrum measured after
an activation of naturally composed Hg of about 12 hours with
$E_{\rm max} = 9900\ {\rm keV}$.

\begin{table*} \caption{Properties of the reactions
$^{196,198,204}$Hg($\gamma$,n) and the following decays of the
produced unstable isotopes $^{195,197,203}$Hg. Additionally, the
reactions $^{198}$Hg($\gamma$,n)$^{197m}$Hg($\gamma$) and
$^{199}$Hg($\gamma$,$\gamma '$)$^{199m}$Hg($\gamma$) are listed.
Data taken from \cite{ensdf}. \label{tab:propHg}}
\begin{ruledtabular} \begin{tabular}{cccccc} seed       & $S_{\rm
n}$ / keV & product     & decay: $T_{1/2}$ & $E_{\gamma}$ / keV &
$I_{\gamma}$ / \%\\ \hline &                   &             &
& 61.46~$\pm$~0.03   & 6.19~$\pm$~0.76 \\ $^{196}$Hg &
8839~$\pm$~50     & $^{195}$Hg  & $\epsilon$: (9.9~$\pm$~0.5)\ h &
779.80~$\pm$~0.05  & 6.8~$\pm$~0.7 \\ &                   &
&                  & 1111.04~$\pm$~0.10 & 1.44~$\pm$~0.20 \\
\hline &                   &             &                  &
77.351~$\pm$~0.002 & 18.7~$\pm$~0.4 \\ $^{198}$Hg & 8484~$\pm$~3
& $^{197}$Hg  & $\epsilon$: (64.14~$\pm$~0.05)\ h &
191.36~$\pm$~0.02  & 0.63~$\pm$~0.02 \\ &                   &
&                  & 268.71~$\pm$~0.03  & 0.039~$\pm$~0.002 \\
\hline $^{204}$Hg & 7495~$\pm$~2      & $^{203}$Hg  & $\beta^{-}$:
(46.61~$\pm$~0.02)\ d & 279.197~$\pm$~0.001& 81.46~$\pm$~0.13 \\
\hline\hline $^{198}$Hg & $\approx 8962$    & $^{197m}$Hg &
$\gamma$: (23.8~$\pm$~0.1)\ h & 133.08~$\pm$~0.05  &
33.48~$\pm$~0.26 \\ \hline &                   &             &
& 158.3~$\pm$~0.1    & 52.3~$\pm$~1.0 \\
\raisebox{2ex}[0cm][0cm]{$^{199}$Hg} &
\raisebox{2.5ex}[0cm][0cm]{---} &
\raisebox{2ex}[0cm][0cm]{$^{199m}$Hg} &
\raisebox{2ex}[0cm][0cm]{$\gamma$: (42.6~$\pm$~0.2)\ min} &
374.1~$\pm$~0.1    & 13.8~$\pm$~1.1 \\
\end{tabular} \end{ruledtabular} \end{table*}

\subsubsection{$^{196}$Hg\label{subsubsec:196hg}} The threshold of
the reaction $^{196}$Hg($\gamma$,n) is $S_{\rm n} = (8839\pm 50)\
{\rm keV}$. Therefore, a considerable ($\gamma$,n) yield could be
obtained only in the activation measurements with $E_{\rm max} =
9450$\ keV and 9900\ keV. The two $\gamma$ lines at $E_{\gamma} =
779.80$\ and 1111.04\ keV were chosen for the analysis.

Both lines occur during the de-excitation of higher lying levels
to the first excited level of $^{195}$Au at $E(J^{\pi} = 1/2^{+})
= 61.4\ {\rm keV}$. This level decays to the ground-state with a
half-life of $T_{1/2} = 3.0\ {\rm ns}$, i.e.\ sumlines have to be
taken into account. This leads to a correction of less than 1\% of
the yield in the single peaks.

Furthermore, the $\gamma$ decay can be observed in coincidence
with the x-rays emitted after the electron capture decay of
$^{195}$Hg. Additional corrections of about 2\% -- 3\% were
necessary for both analyzed $\gamma$ lines.

Due to the fact that only the yields of two irradiation energies
$E_{\rm max}$ could be analyzed the superposition method was not
applicable and the conventional method had to be used. The
assumption of a pure $s$ wave emission leading to the exponent $k
= 0.5$ in Eq.~\ref{eq:sigma} is very likely for the reaction
$^{196}$Hg($\gamma$,n). Table~\ref{tab:196hg} shows the derived
cross section normalization $\sigma_{0}$ for the two $\gamma$
lines at the different activation energies $E_{\rm max}$. All four
values agree within the errors and, consequently, the weighted
mean  $\sigma_{0} = (283 \pm 47)\ {\rm mb}$ has been used to
calculate a ground-state reaction rate of $\lambda_{\rm conv} =
(0.42 \pm 0.07)\ {\rm s}^{-1}$ at $T = 2.5 \times 10^{9}\ {\rm
K}$.

\begin{table} \caption{Results for the normalization $\sigma_{0}$
of Eq.~\ref{eq:sigma} for the reaction $^{196}$Hg($\gamma$,n).
\label{tab:196hg}} \begin{ruledtabular} \begin{tabular}{ccc}
$E_{\rm max}$ / keV & $E_{\gamma}$ / keV & $\sigma_{0}$ / mb \\
\hline & 779.80             & 298~$\pm$~64 \\
\raisebox{2ex}[0cm][0cm]{9450} & 1111.04 & 266~$\pm$~63 \\ &
779.80             & 298~$\pm$~58 \\
\raisebox{2ex}[0cm][0cm]{9900} & 1111.04 & 267~$\pm$~58 \\ \hline
\multicolumn{3}{c}{weighted mean $\langle \sigma_{0} \rangle$:
(283~$\pm$~47)\ mb} \end{tabular} \end{ruledtabular} \end{table}

\subsubsection{$^{198}$Hg\label{subsubsec:198hg}} $^{197}$Hg is
produced in the reaction $^{198}$Hg($\gamma$,n) if the photon
energy exceeds the threshold $S_{\rm n} = (8484 \pm 3)\ {\rm
keV}$. $^{197}$Hg decays by electron capture to $^{197}$Au. The
most prominent $\gamma$ lines emitted during this decay have
energies of $E_{\gamma} = 191.4$\ and 268.7\ keV. Both lines stem
from the decay of a level at $E(J^{\pi} = 3/2^{+}) = 268.7\ {\rm
keV}$ so that the measured yields had to be corrected regarding
summing effects.

The direct decay to the ground-state occurs only with rather low
probabilities (see Table~\ref{tab:propHg}) so that the correction
of the yield for $E_{\gamma} = 191.4\ {\rm keV}$ is in the order
of 1\% whereas the yield for $E_{\gamma} = 268.7\ {\rm keV}$ has
to be corrected by about 10\% . Another minor correction of both
lines are summing effects from x-rays emitted after the electron
capture of $^{197}$Hg. The resulting yields had to be increased by
about 1\%.

The three spectra which could be analyzed were sufficient to use
the superposition method. The uncertainty of the approximation was
in the order of 10\%. The resulting reaction rate at $T = 2.5
\times 10^{9}\ {\rm K}$, $\lambda_{\rm super} = (2.0 \pm 0.3)\
{\rm s}^{-1}$, is in perfect agreement with the reaction rate that
is derived by the conventional method assuming pure $s$ wave
decay: $\lambda_{\rm conv} = (2.0 \pm 0.3)\ {\rm s}^{-1}$. Thus,
the assumed $s$ wave energy dependence of the ($\gamma$,n) cross
section seems to be correct in the observed energy range from
$S_{\rm n}$ to 9900\ keV.

\begin{table} \caption{Results for the normalization $\sigma_{0}$
of Eq.~\ref{eq:sigma} for the reaction $^{198}$Hg($\gamma$,n).
\label{tab:198hg}} \begin{ruledtabular} \begin{tabular}{ccc}
$E_{\rm max}$ / keV & $E_{\gamma}$ / keV & $\sigma_{0}$ / mb \\
\hline & 191                & 294~$\pm$~60 \\
\raisebox{2ex}[0cm][0cm]{9000} & 268     & 277~$\pm$~59 \\ & 191
& 268~$\pm$~46 \\ \raisebox{2ex}[0cm][0cm]{9450} & 268     &
264~$\pm$~48 \\ & 191                & 264~$\pm$~43 \\
\raisebox{2ex}[0cm][0cm]{9900} & 268     & 291~$\pm$~49 \\ \hline
\multicolumn{3}{c}{weighted mean $\langle \sigma_{0} \rangle$:
(278~$\pm$~37)\ mb} \end{tabular} \end{ruledtabular} \end{table}

During the activation an isomeric state in $^{197}$Hg was
populated, too. This state is located at $E = 298.9\ {\rm keV}$
and has spin and parity $J^{\pi} = 13/2^{+}$. It decays with a
half-life $T_{1/2} = (23.8 \pm 0.1)\ {\rm h}$ by electron capture
to $^{197}$Au with a branching of 8.6\% . The state is supposed to
be populated in the ($\gamma$,n) reaction via an intermediate
$J^{\pi} = 9/2^{+}$ state at $E = 478\ {\rm keV}$ \cite{raus00}.
Therefore, the effective ($\gamma$,n) reaction threshold for
isomer population is at about 9\ MeV, and we could neither apply
the superposition method nor the conventional method.

\subsubsection{$^{204}$Hg\label{subsubsec:204hg}} The threshold
for the reaction $^{204}$Hg($\gamma$,n) is located at $S_{\rm n} =
(7495 \pm 2)\ {\rm keV}$. Only one level in $^{203}$Tl is
populated in the $\beta^{-}$ decay of the produced $^{203}$Hg
nuclei. Because this level decays only to the ground-state by
$\gamma$ ray emission with $E_{\gamma} = 279.2\ {\rm keV}$, the
yield need not be corrected for summing effects.

The low reaction threshold has the advantage that the decay line
could be analyzed in all activation spectra down to $E_{\rm max} =
8325\ {\rm keV}$. This limit is given by the determination of the
absolute photon intensity by our calibration standard $^{197}$Au
($S_{\rm n} = 8071\ {\rm keV}$).

This limit is a few hundred keV above the threshold of $^{204}$Hg,
thus, the approximation of the Planck distribution is only
possible with large errors of about 20\% and yields a reaction
rate of $\lambda_{\rm super} = (57 \pm 9)\ {\rm s}^{-1}$. To apply
the conventional method the energy dependence of the ($\gamma$,n)
cross section of Eq.~\ref{eq:sigma} is not valid in the observed
energy range because the distance to the threshold is too large.
One can somehow balance this fact by fitting the exponent to the
data and thereby averaging over a broader energy range.

Table~\ref{tab:204hg} lists the results for the normalization
factors $\sigma_{0}$ derived with a fitted exponent of $k = 0.85$.
The reaction rate calculated with the parameters $k = 0.85$ and
$\langle \sigma_{0} \rangle = (303 \pm 40)\ {\rm mb}$,
$\lambda_{\rm conv} = (58 \pm 8)\ {\rm s}^{-1}$, is in excellent
agreement with the one derived by the superposition method.

\begin{table} \caption{\label{tab:204hg} Results for the
normalization $\sigma_{0}$ for the reaction
$^{204}$Hg($\gamma$,n). A fitted exponent of $k = 0.85$ was used
for the threshold behaviour (see text).} \begin{ruledtabular}
\begin{tabular}{ccc} $E_{\rm max}$ / keV & $E_{\gamma}$ / keV &
$k = 0.85$: $\sigma_{0}$ / mb \\ \hline 8325 & 279.2 &
314~$\pm$~54 \\ 8550 & 279.2 & 277~$\pm$~46 \\ 9000 & 279.2 &
338~$\pm$~54 \\ 9450 & 279.2 & 289~$\pm$~46 \\ 9900 & 279.2 &
307~$\pm$~47 \\ \hline \multicolumn{3}{c}{weighted mean $\langle
\sigma_{0} \rangle$: (303~$\pm$~40)\ mb} \end{tabular}
\end{ruledtabular} \end{table}

\subsection{$^{204}$Pb\label{subsec:Pb}} Pb is the heaviest
element that can be produced in $s$ process nucleosynthesis. The
most proton rich stable isotope is $^{204}$Pb, a $s$-only isotope
that is shielded against the $r$ process flux by $^{204}$Hg.
Therefore, $^{204}$Pb is besides U and Th one of the heaviest
starting points for the $p$ process network calculations on the
proton rich side of the valley of stability.

The determination of the ($\gamma$,n) reaction rate of $^{204}$Pb
has been realized with naturally composed target material. Only
the reactions $^{204}$Pb($\gamma$,n) and $^{206}$Pb($\gamma$,n)
produce unstable isotopes and, additionally, no $\gamma$ rays are
emitted during the electron capture decay of $^{205}$Pb. Thus,
only the $\gamma$ rays following the electron capture decay of
$^{203}$Pb are observable in the activation spectra.

\begin{figure}
\includegraphics[width=\columnwidth]{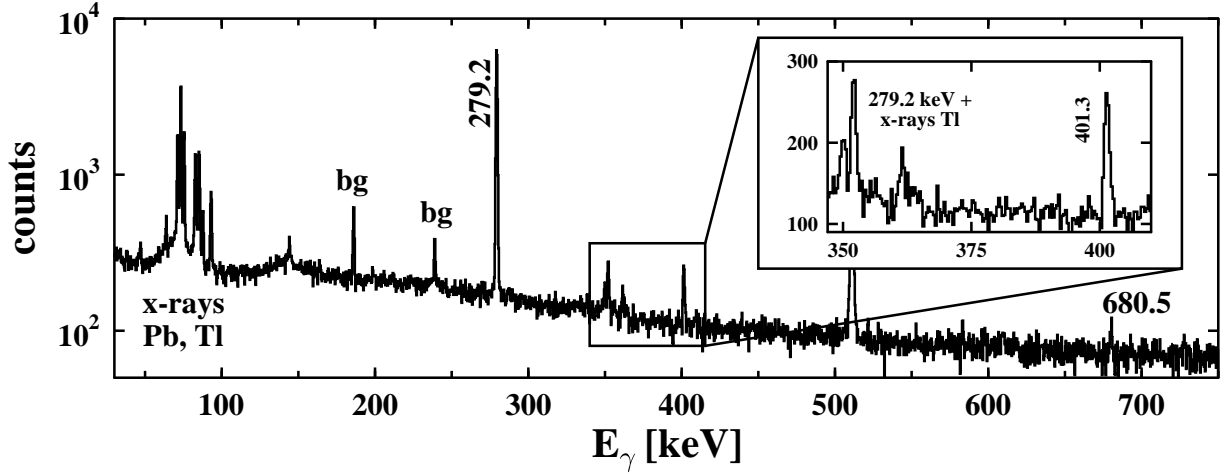} \caption{\label{fig:Pb}
Typical $\gamma$ spectrum after photoactivation of naturally
composed Pb. The spectrum was measured after an activation of
about 24\ hours with $E_{\rm max} = 9900\ {\rm keV}$. The $\gamma$
lines emitted during the decay of $^{203}$Pb are indicated by
their energy $E_{\gamma}$ while background lines are marked by
``bg''. The inset shows the sum of the x-rays of $^{203}$Tl and
the most prominent $\gamma$ line with $E_{\gamma} = 279.2\ {\rm
keV}$.} \end{figure}

The reaction threshold of $^{204}$Pb($\gamma$,n) is $S_{\rm n} =
(8395 \pm 6)\ {\rm keV}$. Thus, the decay lines were observable
after irradiations with $E_{\rm max} = 8775$, 9000, 9450, and
9900\ keV. Figure~\ref{fig:Pb} shows a typical spectrum that was
measured after an irradiation of about 24 hours with $E_{\rm max}
= 9900\ {\rm keV}$.

The half-life of  $^{203}$Pb is $T_{1/2} = (51.87 \pm 0.01)\ {\rm
h}$. During the electron capture decay two levels at $E(J^{\pi} =
3/2^{+}) = 279.2\ {\rm keV}$ and $E(J^{\pi} = 5/2^{+}) = 680.5\
{\rm keV}$ are populated. The latter decays with a probability of
82\% by $\gamma$ emission to the former one. Therefore, three
peaks in the activation spectra at $E_{\gamma} = 279.2$, 401.3,
and 680.5\ keV are observable. The line at 680.5\ keV is composed
by the direct decay of the excited level to the ground-state and
the coincident measurement of the two $\gamma$ rays emitted in a
cascade.

The correction for the yield at $E_{\gamma} = 401.3\ {\rm keV}$
was about 7\% whereas the yield of $E_{\gamma} = 279.2\ {\rm keV}$
has to be corrected by about 0.2\% . A further correction stemming
from coincident measurement of the $\gamma$ rays and the x-rays of
the electron capture decay was taken into account with about 6\% .

The reaction rate derived at $T = 2.5 \times 10^{9}\ {\rm K}$ by
the superposition method is $\lambda_{\rm super} = (1.87 \pm
0.32)\ {\rm s}^{-1}$. If the conventional method is used one has
to consider that an $s$ wave emission is only possible to an
excited level at $E(J^{\pi} = 1/2^{-}) = 125.6\ {\rm keV}$ in
$^{203}$Pb. Thus, an effective reaction threshold of $S_{\rm
n}^{\rm eff} = (8395 + 125.6)\ {\rm keV}$ has to be used if the
energy dependence of the cross section is parameterized as
described in Eq.~\ref{eq:sigma}.

Table~\ref{tab:204pb} summarizes the results for the parameter
$\sigma_{0}$ if  the effective reaction threshold is used. From
the weighted mean a ground-state reaction rate $\lambda_{\rm conv}
= (1.56 \pm 0.25)\ {\rm s}^{-1}$ at $T = 2.5 \times 10^{9}\ {\rm
K}$ is derived which agrees with the result from the superposition
method.

\begin{table} \caption{\label{tab:204pb} Results for the
normalization $\sigma_{0}$ of Eq.~\ref{eq:sigma} for the reaction
$^{204}$Pb($\gamma$,n). An effective threshold of $S_{\rm n}^{\rm
eff} = 8520.6\ {\rm keV}$ was used to calculate $\sigma_{0}^{\rm
eff}$.} \begin{ruledtabular} \begin{tabular}{ccc} $E_{\rm max}$ /
keV & $E_{\gamma}$ / keV & $\sigma_{0}^{\rm eff}$ / mb \\ \hline
8775                           & 279.2   & 285~$\pm$~73 \\ 9000
& 279.2   & 414~$\pm$~74 \\ & 279.2   & 233~$\pm$~34 \\
\raisebox{2ex}[0cm][0cm]{9450} & 401.3   & 257~$\pm$~46 \\ & 279.2
& 231~$\pm$~31 \\ \raisebox{2ex}[0cm][0cm]{9900} & 401.3   &
226~$\pm$~34 \\ \hline \multicolumn{2}{r}{weighted mean $\langle
\sigma_{0} \rangle$: (250~$\pm$~40)\ mb} \end{tabular}
\end{ruledtabular} \end{table}

\section{Comparison to theoretical predictions\label{sec:theory}}
The aim of our measurements is to test the validity of theoretical
predictions of reaction rates at the proton rich side of the
valley of stability. Besides the results reported in this paper
the reaction rates of platinum isotopes \cite{vogt01b} and Au
\cite{vogt02} were already determined so that a systematic survey
in this region is now complete.

Most of the predictions used for $p$ process network calculations
are derived in the Hauser-Feshbach statistical model. The
different results rely on the different treatment of the input
values e.g.\ ground--state properties,
level densities, optical potentials and $\gamma$ ray strength functions.
Some codes use global parameterizations like
NON--SMOKER \cite{raus00} to become as reliable as possible for
unstable and exotic nuclei for which no experimental data is
available. Here the masses are taken from
experiment or the macroscopic FRDM model with
microscopic corrections or the pure microscopic
ETFSI--Q approach. The nuclear level densities 
are derived from a phenomenological Fermi-gas
formalism with microscopic corrections and paring
corrections extracted from the above mentioned
mass models. The optical potentials and the
E1--strength function are derived from global
phenomenological descriptions.
These approaches using global parameterizations
accept the fact that some measured
rates are only calculated within an error range of about 25\% --
30\% \cite{raus97,bao00}. For the NON--SMOKER results shown
in table \ref{tab:comp} experimental masses
and the FRDM mass model haven been chosen.

Other approaches use experimental data if available and mainly
global microscopic or semi-microscopic 
inputs to reproduce the measured
rates very accurately and start their predictions from this basis.
An example is the code MOST \cite{gori98}. 
Here the masses come from experiment or a microscopic
Hartree--Fock--Bogolyubov (HFB--2) model. Nuclear deformations,
pairing properties and the single--particle level
schemes are derived from the HFB--2 approach.
The nuclear level densities stem from a microscopic
statistical model including deformation and pairing effects.
The nucleon--nucleus optical potential is based on a
semi-microscopic Brueckner--Hartree--Fock theory, the 
alpha--nucleus potential stems form a phenomenological
double folding description. Finally the E1--strength function
is from a microscopic QRPA calculation. All references
can be found in the review articel \cite{arno03}.

Our experimental
results are compared to the results of both codes in
Table~\ref{tab:comp}. It can be seen that the agreement between
the experimental data and both predictions is reasonable.

\begin{table} \caption{\label{tab:comp} Comparison of measured and
predicted reaction rates at $T = 2.5 \times 10^{9}\ {\rm K}$. All
values are ground-state reaction rates, thus, the thermalization
of the target nucleus under $p$ process conditions is not taken
into account. Reaction rates predicted with the code NON--SMOKER
\cite{raus00} are indicated by the subscript ``NONS'' and
predictions with the code MOST \cite{gori98} by ``MOST''.}
\begin{ruledtabular} \begin{tabular}{ccccc} isotope    &
$\lambda_{\rm exp}^{\rm super}$ / s$^{-1}$ & $\lambda_{\rm
exp}^{\rm conv}$ / s$^{-1}$ & $\lambda_{\rm theo}^{\rm NONS}$ /
s$^{-1}$ & $\lambda_{\rm theo}^{\rm MOST}$ / s$^{-1}$ \\ \hline
$^{190}$Pt & ---           & 0.4~$\pm$~0.2   & 0.18 & 0.29 \\
$^{192}$Pt & 0.5~$\pm$~0.2 & 0.4~$\pm$~0.1   & 0.58 & 0.56 \\
$^{198}$Pt & 87~$\pm$~21   & 73~$\pm$~17     & 50   & 110  \\
$^{197}$Au & 6.2~$\pm$~0.8 & 5.8~$\pm$~0.8   & 4.8  & 5.6  \\
$^{196}$Hg & ---           & 0.42~$\pm$~0.07 & 0.32 & 0.58 \\
$^{198}$Hg & 2.0~$\pm$~0.3 & 2.0~$\pm$~0.3   & 1.4  & 2.1  \\
$^{204}$Hg & 57~$\pm$~9    & 58~$\pm$~8      & 73   & 170  \\
$^{204}$Pb & 1.9~$\pm$~0.3 & 1.6~$\pm$~0.3   & 1.5  & 3.0
\end{tabular} \end{ruledtabular} \end{table}

\begin{figure}
\includegraphics[width=\columnwidth]{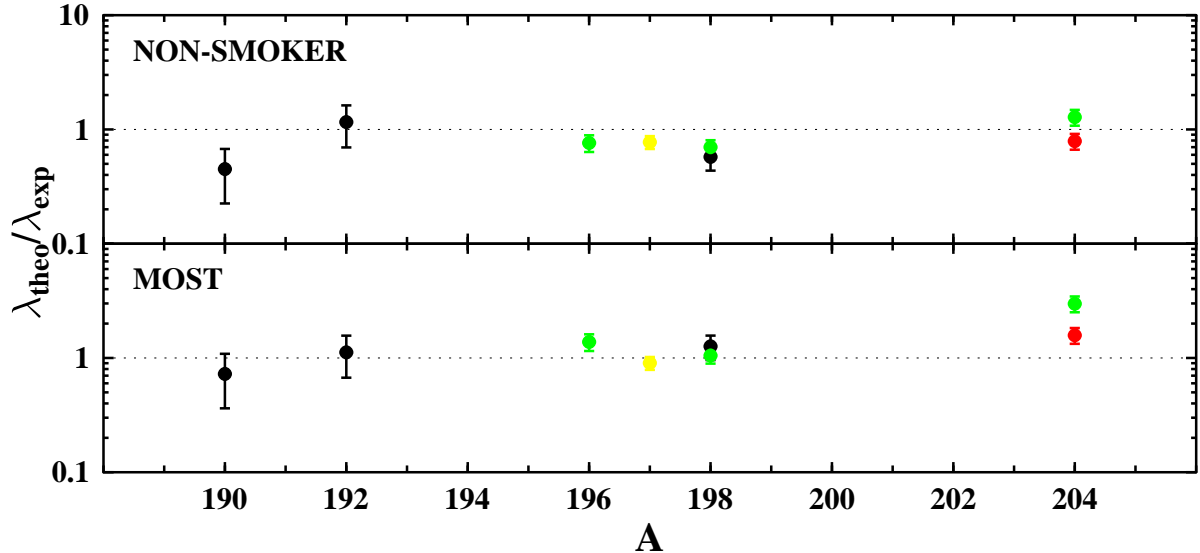}
\caption{\label{fig:verglmass} (Color online) Ratios between
theoretically predicted and experimentally derived reaction rates.
The upper panel uses the predictions by the code NON--SMOKER
\cite{raus00} whereas in the lower panel the values predicted by
the code MOST \cite{gori98} are used. The black dots correspond to
platinum isotopes \cite{vogt01b}, the yellow one to $^{197}$Au
\cite{vogt02}, the green ones to Hg isotopes, and the red one to
$^{204}$Pb.} \end{figure}

Figure~\ref{fig:verglmass} shows the ratio between the theoretical
and experimental values. The experimental reaction rates are
derived by the superposition method if available. No systematic
deviation can be seen in the observed mass region neither for the
predictions of the code NON--SMOKER (upper panel) mostly using
phenomenogical models for the input values nor for the code
MOST (lower panel) relying mostly on microscopic models. 
This can be interpreted as a good base for
predictions of reaction rates of unstable or exotic nuclei in this
mass region. Nevertheless, the deviations for some nuclei range in
both cases to factors of up to 2. Thus, further studies seem to be
required to improve the nuclear physics input of the codes.

\section{Summary and outlook\label{sec:summary}} Theoretical
predictions of ($\gamma$,n) reaction rates are mandatory as an
input for $p$ process network calculations. The validity of the
predictions should be tested by comparing the predicted values to
experimental ones. Unfortunately, most of the experimental data
was measured around the Giant Dipole Resonance, hence, far off the
astrophysically relevant energy region.

We have shown that it is possible to determine ground-state
reaction rates using the photoactivation technique without any
model dependency by a superposition of bremsstrahlung spectra.
This method has been used in the mass region $Z \geq 78$ to study
the theoretical predictions systematically. No systematic
deviation between the experimental rates and those predicted by
the codes NON--SMOKER and MOST have been found.

Due to the high temperatures reached in stellar environments
excited levels are populated by thermalization resulting in
reaction rates increased by several orders of magnitude. The
so-called stellar enhancement factor $\lambda^{*} / \lambda^{\rm
gs}$ cannot be measured and is fully based on theoretical
predictions. The effect is sensitive to level densities and
underlying nuclear structure, thus, being a further source of
systematic errors.

The mass region $A \approx 100$ is of special interest because
this is the border region between the different processes
responsible for the production of the $p$ nuclei. A systematic
study of the validity of theoretical predictions of ($\gamma$,n)
and ($\gamma$,$\alpha$) reaction rates would be desirable to
improve the understanding of $p$ process nucleosynthesis.

\begin{acknowledgments} We thank the other members of our group
especially M.\ Babilon, W.\ Bayer, K.\ Lindenberg, D.\ Savran, and
S.\ Volz for their support during the beamtime. We also thank P.\
Mohr and T.\ Rauscher for valuable discussions. This work was
supported by the Deutsche Forschungsgemeinschaft under contract
SFB 634.  \end{acknowledgments}

\bibliography{/userx/users/kerstin/bibtex/bibtex/english}

\end{document}


\preprint{}

\title{Erratum: Systematic study of ($\gamma$,n) reaction rates 
for $Z \geq 78$ isotopes [Phys.\ Rev.\ C {\bf 70}, 035802 (2004)]}


\author{K.\ Sonnabend}
\author{K.\ Vogt}
\author{D.\ Galaviz}
\author{S.\ M\"uller}
\author{A.\ Zilges}
\email{zilges@ikp.tu-darmstadt.de}
\affiliation{Institut f\"ur Kernphysik, Technische Universit\"at Darmstadt, Schlossgartenstr.\ 9, D--64289 Darmstadt, Germany}


\date{\today}

\pacs{
25.20.-x, 
26.30.+k, 
27.80.+w  
}

\maketitle


A typographical error occured in Table~I of the original paper \cite{sonn04b}. The natural abundance of $^{196}$Hg was cited incorrectly. The correct value is (0.15~$\pm$~0.01)\%.

The unit of the photon flux $N_{\gamma}$ plotted as the ordinate in Figure~2 of the original paper \cite{sonn04b} is incomplete. The unit should read ${\rm keV}^{-1}{\rm cm}^{-2}{\rm s}^{-1}$.

The half-lives of the Hg isotopes have been remeasured by \cite{lind01}. These values were used in the analysis if their accuracy was significantly improved compared to the values given in \cite{ensdf}. Thus, column four of Table~II of the original paper \cite{sonn04b} has been corrected (see Table~\ref{tab:propHg}).
\begin{table*} 
\caption{Properties of the reactions $^{196,198,204}$Hg($\gamma$,n) and the following decays of the produced unstable isotopes $^{195,197,203}$Hg. Additionally, the reactions $^{198}$Hg($\gamma$,n)$^{197m}$Hg($\gamma$) and $^{199}$Hg($\gamma$,$\gamma '$)$^{199m}$Hg($\gamma$) are listed. Half-lives of $^{195,197,199m}$Hg were taken from \cite{lind01}, other data was taken from \cite{ensdf}. \label{tab:propHg}}
\begin{ruledtabular}
\begin{tabular}{cccccc} 
Seed       & $S_{\rm n}$ / keV & Product     & Decay: $T_{1/2}$ & $E_{\gamma}$ / keV & $I_{\gamma}$ / \%\\ \hline
$^{196}$Hg & 8839~$\pm$~50     & $^{195}$Hg  &  $\epsilon$: (10.53~$\pm$~0.03)\ h & 61.46~$\pm$~0.03   & 6.19~$\pm$~0.76 \\ 
           &                   &             &                                 & 779.80~$\pm$~0.05  & 6.8~$\pm$~0.7 \\
           &                   &             &                                 & 1111.04~$\pm$~0.10 & 1.44~$\pm$~0.20 \\
$^{198}$Hg & 8484~$\pm$~3      & $^{197}$Hg  & $\epsilon$: (64.94~$\pm$~0.07)\ h & 77.351~$\pm$~0.002 & 18.7~$\pm$~0.4 \\ 
           &                   &             &                                   & 191.36~$\pm$~0.02  & 0.63~$\pm$~0.02 \\
           &                   &             &                                   & 268.71~$\pm$~0.03  & 0.039~$\pm$~0.002 \\
$^{204}$Hg & 7495~$\pm$~2      & $^{203}$Hg  & $\beta^{-}$: (46.61~$\pm$~0.02)\ d & 279.197~$\pm$~0.001& 81.46~$\pm$~0.13 \\
$^{198}$Hg & $\approx 8962$    & $^{197m}$Hg & $\gamma$: (23.8~$\pm$~0.1)\ h & 133.08~$\pm$~0.05  & 33.48~$\pm$~0.26 \\
$^{199}$Hg & ---               & $^{199m}$Hg & $\gamma$: (42.67~$\pm$~0.09)\ min & 158.3~$\pm$~0.1    & 52.3~$\pm$~1.0 \\
           &                   &             &                                 & 374.1~$\pm$~0.1    & 13.8~$\pm$~1.1 \\
\end{tabular} 
\end{ruledtabular} 
\end{table*}

During the preparation of this erratum the error propagation was checked, too. The approximation of the Planck distribution is possible with an overall mean deviation of about 5 - 10\%. This uncertainty has been included in the statistical errors presented in Table~VII of the original paper \cite{sonn04b}. However, we found that a systematical uncertainty of about 20\% was mentioned in the text in the special case of $^{204}$Hg but it was not included in the results presented in Table~VII of the original paper \cite{sonn04b}. This systematic uncertainty results from an underestimation of the photon flux, thus, the reaction rate is underestimated, too. The corrected reaction rate is $\lambda_{\rm exp}^{\rm super}(^{204}{\rm Hg}) = (57 \pm 9 + 12)\ {\rm s}^{-1}$.

Furthermore, the results of the ground-state reaction rates for the Pt isotopes have changed compared to previous publications \cite{mohr00b,vogt01b} due to an improved analysis. The differences are described in detail in \cite{vogt03b}.


The final results for the ground-state reaction rates presented in \cite{sonn04b} are not influenced by these corrections.

Finally, we want to stress that the neutron density during $r$-process nucleosynthesis is well above $10^{20}\ {\rm cm}^{-3}$ (see e.g.\ \cite{krat93}).

\begin{acknowledgments} 
We thank P.\ Mohr for his careful reading of the original paper. This work was supported by the Deutsche Forschungsgemeinschaft under contract SFB 634.  
\end{acknowledgments}

\bibliography{/userx/users/kerstin/pub/bibtex/bibtex/english}